\documentclass[10pt,a4paper,twoside]{article}
\title{\bf A simple direct quantum model which, with no random phase assumptions and with arbitrary initial conditions, evolves to the Boltzman distribution}
\author{Michael J. Caola\\6 Normanton Rd., Bristol, BS8 2TY, UK \\ caolam@blueyonder.co.uk}
\usepackage{graphics}
\usepackage{amsmath}
\usepackage{amssymb}
\begin{document}

\maketitle 
\begin{abstract}
We consider M systems (each an electron in a long square cylinder) uniformly arranged on a ring and with  Coulomb interactions. Exact straightforward numerical time-dependent perturbation calculation of a single N-level ($\lesssim 7$) system, with no (random) phase assumptions, system show a Boltzman distribution. We exploit the physical ring symmetry and develop several hierarchical physical equation set so of increasing generality and (computation) speed. Given the impressive history of theoretical quantum-mehanical statistical mechanics, our results might seem surprising, but we observe that accurate calculation of correct physical equations should mimic Nature.    

\noindent
\\
\\
                                                   
\end{abstract}

\section{Introduction}
An important part of Statistical Mechanics is to explain how a physical system can tend to an equilibrium state, usually based on its energy structure.
Most analysis, both classical and quantum-mechanical, invokes probability assumptions which in the latter are additional to those of the basic non-relativistic Schrodinger wave-function[,,]. We treat the quantum case and in particular wish to understand how a quantum system can (in time) tend to a state
\begin{equation}  
\psi=\sum_{n=1}^N c_n \psi_n  
\end{equation}
whose (eigen)states $ \psi_n$ and (eigen)energies $E_n$ are  defined by Hamiltonian $H_0$: $H_0 \psi_n=E_n \psi_n$, $\omega_n=E_n/\hbar$, $\Psi_n(\bf r,t)=\psi_n(\bf r) e^{-i\omega_n t}$. The expansion coefficient $c_n=c_n(t\rightarrow\infty)$ gives the probability $w_n$ that in equilibrium  the system is in state $\psi_n$:
\begin{equation}
w_n=|c_n|^2=e^{-\beta E_n}/\sum_ne^{-\beta E_n}
\end{equation}
This (2) is the Boltzman distribution where $\beta=1/kT$, $T=$ temperature and $k$ is the Boltzmann constant.

\section{Analysis}
\subsection{Model geometry}
Our model is $M$  identical systems uniformly on a ring radius $R$ and normal $x$. The model is isolated (from the rest of the universe). Each system is an electron in a "matchstick" box which is a long $a$ square $b\ll a$ prism, see fig.1.  The physical environment each system $m=1..M$ is that of its $M-1$ neighbours, and inspection/symmetry shows that this is the same for all $M$ systems: every system $m=1..M$ is in the 'same heat bath'. This 'same heat bath' is essentially a $2D$ model with {\it finite} $M$, and it would seem that in $1D$ or $3D$ only $M=\infty$ is possible.
\par
The energy of an electron in an infinitely deep potential rectangular box, sides $(a_x,a_y,a_z)$, is
\begin{equation} 
E_{n_x,n_y,n_z}=\sqrt\frac{8}{a_x a_y a_z}\left(\frac{n_x^2}{a_x^2}+\frac{n_y^2}{a_y^2}+\frac{n_z^2}{a_z^2}\right)
\end{equation} 
and eignfunctions, $\omega=E/\hbar$, 
\begin{equation}
\Psi_{n_x,n_y,n_z}(x,y,z,t)=\psi_{n_x,n_y,n_z}(x,y,z)e^{-i\omega_{n_x,n_y,n_z}t}
\end{equation} 
\begin{equation}
\psi_{n_x,n_y,n_z}(x,y,z)=\sqrt\frac{8}{a_x a_y a_z}\left(\sin\frac{\pi n_x}{a_x}x\sin\frac{\pi n_y}{a_y}y\sin\frac{\pi n_z}{a_z}z\right)
\end{equation}
\begin{equation}
\equiv\psi_{n_x}(x)\psi_{n_y}(y)\psi_{n_z}(z)
\end{equation}
Our matchstick system with $\{a,b\ll a\}$ means that energies $(E_{N,1,1}-E_{1,1,1})\ll E_{1,2,1}$: the $N$ lowest energy states enjoy a large energy separation from higher ones and it is a well-established quantum common-place that the effects of a perturbation on the system may be accurately calculated using  these $N$ states only --- we have a $N$-level system. Thus,with $n_x\equiv n$, we henceforth deal only with states $\psi_{n,1,1}$, $n=1..N$
\begin{equation}
\psi_{n}(x)=\psi_{n,1,1}(x,y,z)=\sqrt\frac{8}{a b^2}\left(\sin\frac{\pi n}{a}x\sin\frac{\pi}{b}y\sin\frac{\pi}{b}z\right)
\end{equation} 
The last part of the geometry is, classically stated, the separation $r_{m,m'}$ of electrons at positions $x_m$ in system $m$ from $x_{m'}$ in system $m'$, see Fig.1:
\begin{equation}
r_{m,m'}(x_m,x_{m'})=\sqrt {(x_m-x_{m'})^2+4R^2\sin^2\Bigl(\frac{\pi(m-m')}{M}\Bigr)}
\end{equation}

\subsection{Model perturbations}
If there are no interactions between systems then system $m$ has wave-function (1)
\begin{equation}
\Psi^{m}=\sum_{n=1}^N \bar{c}_n^{m} \Psi_n 
\end{equation}
where probability $\bar{w}_n^{m}=|\bar{c}_n^{m}|^2$ is independent of time $t$, but otherwise arbitrary. In natural fact the electron $e_m$ experiences the Coulomb repulsion $V^m$ of its $M-1$ neighbours $e_{m'\ne m}$:
\begin{equation}
dV^m=\sum_{m'\ne m}^N \frac {e_m de_{m'}}{r_{mm'}}
\end{equation}
\par
In Eq.(10) we treat $e_m=e$ classically as a point-charge electron, and $e_{m'}$ quantum mechanically: $de_{m'}=e|\psi^{m'}(x_{m'})|^2dx_{m'}$ is the infintesimal charge within $dx_{m'}$ of system $m'$, so with $r_{mm'}$ of Eq.(8), 

\begin{equation}
V^m=\sum_{m'\ne m}^N e_m \int_0^a \frac { de_{m'}}{r_{mm'}}=e^2\sum_{m'\ne m}^N   \int_0^a \frac { |\psi^{m'}(x_{m'})|^2dx_{m'}}{\sqrt {(x_m-x_{m'})^2+4R^2\sin^2\Bigl(\frac{\pi(m-m')}{M}\Bigr)}}
\end{equation}
With, c.f. Eq.(1,9), 
\begin{equation}
\Psi^{m'}=\sum_{n=1}^N c_n^{m'} \Psi_n 
\end{equation}
we finally have that the time-dependent perturbation on system $m=1..M$ caused by its $M-1$ neighbour systems is
\begin{equation}
V^m(x_m,t)=e^2\sum_{n_{1},n_{2}}^{N}\sum_{m'\ne m}^{M}c_{n_{1}}^{m'}(t) c_{n_{2}}^{m'}(t)^*\int_0^a\frac{\Psi_{n_{1}}^{m'}(x_{m'},t) \Psi_{n_{2}}^{m'}(x_{m'},t)^* dx_{m'}}{\sqrt {(x_m-x_{m'})^2+4R^2\sin^2\Bigl(\frac{\pi(m-m')}{M}\Bigr)}}  
\end{equation}

\subsection{Perturbation calculations}
We summarise nearly 100 year-old time-dependent QM perturbation theory, with general $H_0$, $\psi_n$ and $V(t)$. The exact solution to the TDSE $i\hbar\partial\Psi/\partial t=(H_0+V(t))\Psi$ is $\Psi=\sum_n c_n\Psi_n$, where [L\&L]
\begin{equation}
\dot{c}_n=\frac{1}{i\hbar}\sum_{n'}c_{n'}\int\Psi_{n'}^{*}({\bf r},t)V({\bf r},t)\Psi_{n}({\bf r},t)d{\bf r}.
\end{equation}
With correspondance $(H_0,\psi_n,\Psi_n,c_n,V,d{\bf r})\Rightarrow(H_0^m,\psi_n^m,\Psi_{n,1,1}^m,c_n^m,V^m,dxdydz)$, substitution of Eqns(4,7,8) into Eqn(14) and evaluation of the $\int\int..dydz$ integral give
$$
\dot{c}_n^m(t)=\frac{4e^2a^2}{i\hbar}\sum_{n',n_{1},n_{2}}^N e^{i(\omega_{n}-\omega_{n'}+\omega_{n_{2}}-\omega_{n_{1}})t}c_{n'}^{m}(t)\sum_{m'\ne m}^M c_{n_{1}}^{m'}(t)c_{n_{2}}^{m'}(t)^*
$$
\begin{equation}
\int_0^a\int_0^a\frac{dxdx'\sin(\pi nx/a)\sin(\pi n'x/a)\sin(\pi  n_{1} x'/a)\sin(\pi n_{2}x'/a)}{\sqrt{(x-x')^2+4R^2\sin^2\Bigl(\frac{\pi(m-m')}{M}\Bigr)}}
\end{equation}
 Also, to conventionally manipulate the differential set Eq.(15), we need to replace the integer pair ($m,n$) by the single integer $p$: counting in mixed base ($M,N$) gives $p=m+(n-1)M$. Eq.(15) then becomes
$$
\dot{c}_{p(m,n)}=\frac{4e^2a^2}{i\hbar}\sum_{n',n1,n2}^Nc_{p(m,n')}\sum_{m'\ne m}^M c_{p(m',n_{1})}c_{p(m',n_{2})}{^*} e^{i(\omega_{n}-\omega_{n'}+\omega_{n_{2}}-\omega_{n_{1}})t}
$$
\begin{equation}
\int_0^a\int_0^a\frac{dxdx'\sin(\pi nx/a)\sin(\pi n'x/a)\sin(\pi n_{1}x'/a)\sin(\pi n_{2}x'/a)}{\sqrt{(x-x')^2+4R^2\sin^2(\pi(m-m')/M)}}
\end{equation}
$$
p=p(m,n)=m+(n-1)M\;\;\;\;\;\;\;\; m=1..M \;\;\;\;\;n=1..N \;\;\;\;\;p=1..MN
$$
To solve/use Eq.(16) we must give them initial values $\bar{c}_p=c_p(t=0)$, let them solve until time t, and then examine the probabilities $w_p(t)=|c_p(t)|^2$. A typical choice of initial values is $\bar{c}_p=1/\sqrt N$: all initial probabilities are equal, which incidentally corresponds to a temperature $T=\infty$.
\par
Note that Eq.(15) is non-linear, $ c_{p}c_{p'}c_{p''}*$ occurs on the r.h.s. We are still using the traditional linear Schr\"{o}dinger equation Eq.(14)
and the non-linearity arises from our formulation $\S$2.2 of the perturbation $V(t)$.
\par
If we put $d^m_n(t)=c^m_n(t) e^{ -i \omega_n t}$, use $\dot{c}^m_ne^{-i\omega_n t }=\dot{d}^m_n+i\omega_n d_n^m$ in (16), and revert $d\to c$, we obtain the 'Interaction Representation' (IR):
$$
\dot{c}_{p(m,n)}=-i\omega_n c_{p(m,n)}+\frac{4e^2a^2}{i\hbar}\sum_{n',n1,n2}^Nc_{p(m,n')}\sum_{m'\ne m}^M c_{p(m',n_{1})}c_{p(m',n_{2})}{^*}
$$
\begin{equation}
\int_0^a\int_0^a\frac{dxdx'\sin(\pi nx/a)\sin(\pi n'x/a)\sin(\pi n_{1}x'/a)\sin(\pi n_{2}x'/a)}{\sqrt{(x-x')^2+4R^2\sin^2(\pi(m-m')/M)}},
\end{equation}
which may be useful in numerical evaluation.

\subsection{Symmetry reductions}
We shall see that solutions $w_{p}(t)$ of Eq.(16) are often ``steady-state oscillatory''. We assume that any physical measurent must last a (small) finite time $\Delta t$, so that any measured or inferred physical probability $W_{p}(t)$ is an an average over many ``oscillations'',
\begin{equation}
W_{p}(t)=\frac{1}{\Delta t}\int_{t}^{t+\Delta t}w_{p}(t') dt'
\end{equation}
One might call (16) coarse-graining, a form of RPA: but this occurs at time $t$, and there is no RPA from $0 \to t$ with evolution following (16).

\subsubsection{P1: Finite $M$ individual systems, $\it\bf {c_n^m}$}
This is Eq.(16) which we recall for convenience:
$$
\dot{c}_{p(m,n)}=\frac{4e^2a^2}{i\hbar}\sum_{n',n_{1},n_{2}}^Nc_{p(m,n')}\sum_{m'\ne m}^M c_{p(m',n_{1})}c_{p(m',n_{2})}{^*} e^{i(\omega_{n}-\omega_{n'}+\omega_{n_{2}}-\omega_{n_{1}})t}
$$
\begin{equation}
\int_0^a\int_0^a\frac{dxdx'\sin(\pi nx/a)\sin(\pi n'x/a)\sin(\pi n_{1}x'/a)\sin(\pi n_{2}x'/a)}{\sqrt{(x-x')^2+4R^2\sin^2(\pi(m-m')/M)}}
\end{equation}
$$
p=p(m,n)=m+(n-1)M\;\;\;\;\;\;\;\; m=1..M \;\;\;\;\;n=1..N \;\;\;\;\;p=1..MN.
$$
\subsubsection{P2: Finite $M$ similar systems, $\bf c_n$}
We now exploit the physical symmetry of our model to simplify  the complex amplitude $c_{n}^{m}$. We start all systems with the same initial probability,$w_{n}^{m}(0)=w_{n}(0)$  and assume/postulate that this holds for all future $t$, $w_{n}^{m}(t)=w_{n}(t)$. Since the environment of each site $m=1..M$ is the same and each interacts with the $ M-1$ others, we can expect the probability to be the same; $w_{n}^{m}=w_{n}^{m'}=w_{n}$. With $c_{n}^{m}\equiv r_{n}^{m} e^{i\theta_{n}^{m}}$ (pure math polar form, real $r$ and $\theta$, no phys.), this means that $w_{n}=(r_n^m)^2 \equiv r_n^2$. We postulate that $\theta_n^m =\theta_n(t)+\theta^m(0) $ where $\theta^m$ is time-independent, reflecting the static invariance of our ring geometry. We now have $c_{n}^{m}(t)=r_n(t)e^{i\theta_n(t)}e^{i\theta^m}$ whose substitution in Eq.(15 or 16) eliminates  $e^{i\theta^m}$:
\begin{equation}
c_{n}(t)=r_n(t)e^{i\theta_n(t)}
\end{equation}
We can show that in Eq.(16)  $\sum_{m'\ne m}^{M}\rightarrow \sum_{m'=1}^{M-1}$ (independent of $m$) , we 'normalise' $\int_0^a\rightarrow \int_0^1$, finally giving 
$$
\dot{c}_{n}=\frac{4e^2}{i\hbar} \sum_{n',n_{1},n_{2}}^N c_{n'} c_{n_{1}}c^*_{n_{2}} e^{i(\omega_{n}-\omega_{n'}+\omega_{n_{2}}-\omega_{n_{1}})t}
$$
\begin{equation}
\int_0^1\int_0^1\sum_{m'=1}^{M-1}\frac{dxdx'\sin(\pi nx)\sin(\pi n'x)\sin(\pi n_{1}x')\sin(\pi n_{2}x')}{\sqrt{a^2(x-x')^2+4R^2\sin^2(\pi m'/M)}}
\end{equation}
$$
 m=1..M \;\;\;\;\;n=1..N    
$$

\subsubsection{P3: Infinite $M$ similar systems, \bf $c_n$}
We are interested in a system in an infinite $M\rightarrow \infty$ heat-bath, and the sum
$$
\sum_{m'=1}^{M-1}\frac{1}{\sqrt{a^2(x-x')^2+4R^2\sin^2(\pi m'/M)}}
$$
in Eq.(20) can be approximated by an integral $\int_1^\infty..dm'$ which Mathematica 7 gives as an Elliptic $EK=$
$$
EllipticK\left(\frac{-4R^2}{a^2(x-x')^2}\right).
$$
Eq.(20) then becomes
$$
\dot{c}_{n}=\frac{4e^2}{i\hbar} \sum_{n',n_{1},n_{2}}^N c_{n'} c_{n_{1}}c^*_{n_{2}} e^{i(\omega_{n}-\omega_{n'}+\omega_{n_{2}}-\omega_{n_{1}})t}
$$
\begin{equation}
\int_0^1\int_0^1dxdx'\sin(\pi nx)\sin(\pi n'x)EK\left(\frac{-4R^2}{a^2(x-x')^2}\right)\sin(\pi n_{1}x')\sin(\pi n_{2}x')
\end{equation}
$$
n=1..N
$$

\section{Numerical results}
We have implemented the above analysis in Mathematica 12 to give numerical results, particularly the measured probability
\begin{equation}
W_{p}(t)=\frac{1}{\Delta t}\int_{t}^{t+\Delta t}w_{p}(t') dt'.
\end{equation}
A typical result for initial $\bar{c}_p=c_p(t=0)=1/\sqrt(N)$, $N=3$ and $M=3$ of actual (instaneous) $w_{p}(t)$ and measured (experimental) $W_{p}(t)$ probabilities is shown in Fig.2. Visually, the intantaneous oscillatory chaos is converted to experimentally observed constant results.


Numerically, with energies $E_{n} \sim n^{2}$ the ratio $(E_{1}-E_{2})/(E_{2}-E_{3})$ is 0.6 theoretically; physically we use (2)and
 $Log(w_{n})$, see Fig.2, which give $(0.78-1.10)/(1.10-1.59)=0.65$
\\\\
Similar results are obtained for variations of $N,M,a,R$, the three cases of $\S 2$ and the important initial conditions $c^m_n(t=0)$.

\section{Discussion}
The preceding may be regarded as a useful direct quantitative pedagogic illustration of the quantum {\itshape density matrix} (statistical operator), see Landau \& Lifshitz [1].
\\
To broadly summarize: at $t=0$ our system is in an arbitrary initial condition, whence it evolves as in (16) (with no random phase assumptions) to a Boltzman distribution.
\section{References}
[1] Landau L.and Lifschitz E. (1977), {\itshape Quantum Mechanics} 3e, Pergamon Press.
\end{document}